\documentclass[conference, 10pt]{IEEEtran}
\usepackage{cite}
\usepackage{comment}
\usepackage{amsmath,amssymb,amsfonts}
\usepackage{algorithmic}
\usepackage{graphicx}
\usepackage{textcomp}
\usepackage{xcolor}
\def\BibTeX{{\rm B\kern-.05em{\sc i\kern-.025em b}\kern-.08em
    T\kern-.1667em\lower.7ex\hbox{E}\kern-.125emX}}
\newtheorem{theorem}{\bfseries Theorem}
\newtheorem{lemma}{\bfseries Lemma}
\newtheorem{definition}{\bfseries Definition}

\newtheorem{remark}{\bfseries Remark}

\newcommand{\hX}{\hat{X}}

\newcommand{\pdf}[1]{ p_{#1} }

\newcommand{\p}[1]{p_{X}}
\newcommand{\ph}[1]{p_{\hat{X}}}

\newcommand{\E}[2][]{ 
    \mathbb{E}_{#1} \left[ #2 \right]
}

 \def\dx{\mathrm{d}x}

\begin{document}

\title{Analyzing $\alpha$-divergence in Gaussian Rate-Distortion-Perception Theory}

\author{Martha V. Sourla\IEEEauthorrefmark{1}, Giuseppe Serra\IEEEauthorrefmark{2}, Photios A. Stavrou\IEEEauthorrefmark{2}, Marios Kountouris\IEEEauthorrefmark{2}\IEEEauthorrefmark{3}\\
\IEEEauthorrefmark{1}{School of Electrical \& Computer Engineering, Technical University of Crete, Chania, Greece}\\
\IEEEauthorrefmark{2}{Communication Systems Department, EURECOM, Sophia-Antipolis, France}\\
\IEEEauthorrefmark{3}{Department of Computer Science and Artificial Intelligence, Andalusian Research Institute}\\  
\IEEEauthorrefmark{0}{in Data Science and Computational Intelligence (DaSCI), University of Granada, Spain}\\             
                    
\texttt{msourla@tuc.gr, \{giuseppe.serra, fotios.stavrou\}@eurecom.fr, mariosk@ugr.es}
\vspace{-1mm}}

\maketitle

\begin{abstract}
The problem of estimating the information rate distortion perception function (RDPF), which is a relevant information-theoretic quantity in goal-oriented lossy compression and semantic information reconstruction, is investigated here. Specifically, we study the RDPF tradeoff for Gaussian sources subject to a mean squared error (MSE) distortion and a perception measure that belongs to the family of $\alpha$ divergences. 
Assuming a jointly Gaussian RDPF, which forms a convex optimization problem, we characterize an upper bound for which we find a parametric solution. We show that evaluating the optimal parameters of this parametric solution is equivalent to finding the roots of a reduced exponential polynomial of degree $\alpha$. Additionally, we determine which disjoint sets contain each root, which enables us to evaluate them numerically using the well-known bisection method. Finally, we validate our analytical findings with numerical results and establish connections with existing results.

\end{abstract}


\section{Introduction}
Rate-distortion-perception (RDP) theory, which has led to a surge of research recently, considers the problem of lossy compression under perceptual constraints on the reconstructed samples, generalizing the classical rate-distortion (RD) theory. Its mathematical representation is given by the rate-distortion-perception function (RDPF), whose properties are first studied in \cite{blau:2018,blau:2019,matsumoto:2019}. 
The rationale behind this approach is rooted in the observation that minimizing the distortion does not necessarily result in visually pleasing human perception results. Interestingly, divergence constraints can be interpreted as semantic quality metrics, which quantify the relevance and significance of the reconstructed source from the observer's perspective in goal-oriented semantic communication \cite{kountouris:20} (see also \cite{stavrou:2023}).

The RDPF, like the conventional RDF, is devoid of an analytical solution for generic sources.
However, there are some efforts to obtain closed-form expressions for specific sources under several commonly used metrics related to distortion and perception. Some of these efforts are reported in \cite{blau:2019,serra:2024}, focusing on discrete binary sources under Hamming distance as a distortion measure and total-variation distance as a perception measure, and also assuming continuous Gaussian sources under MSE distortion and either Wasserstein-2 distance, Hellinger distance, Geometric Jensen-Shannon divergence, or (direct and reverse) Kullback-Leibler (KL) divergence perception constraints. {A generic method to compute the RDPF for continuous alphabet sources (not necessarily Gaussian),  when the distribution of the source is constrained to be the same as the distribution of the output, is studied in \cite{serra:2024copula}.}

The key contribution of this paper is a closed-form solution to the RDPF of a Gaussian distributed source subject to MSE distortion and an $\alpha$ divergence perception constraint.
Alpha divergences, which were first proposed by Chernoff \cite{Chernoff:52} and are thoroughly investigated by Amari \cite{Amari:82,Amari:85}, are unique canonical divergences that reside at the intersection of $f$-divergences and Bregman divergences in a manifold of positive measures. Since $\alpha$-divergences are convex on both arguments, the RDPF forms a convex optimization problem. By constraining the reconstruction to be jointly Gaussian with the source, we define the jointly Gaussian RDPF (JG-RDPF) and show that it is an upper bound on the RDPF. Next, we provide a parametric solution to the JG-RDPF, for which we need to find the roots of a reduced exponential polynomial of degree $\alpha$, derived from the perception constraint. We show that the reduced polynomial has a unique global stationary point and that there exist two unique roots that belong to disjoint sets for every value of $\alpha$. The two roots can be approximated numerically using the bisection method. Subsequently, we provide a new parametric closed-form expression for Pearson's divergence ($\alpha = 2$), which is a specific instance of $\alpha$ divergence. We verify our theoretical findings with numerical evaluations.

\par{\textit{Notation}:}
Let $\mathcal{X}$ be a Euclidean space (possibly finite-dimensional), and $\mathbb{B}(\mathcal{X})$ the Borel $\sigma$-algebra on $\mathcal{X}$. A random variable $X$ defined on some probability space ($\Omega, \mathcal{F}, \mathcal{P}$) is a map $X:\Omega\mapsto\mathcal{X}$, where $(\mathcal{X}, \mathbb{B}(\mathcal{X}))$ is a measurable space. Given a continuous and twice differentiable function $f(x)$, the notation $f'(x)$ and $f''(x)$ denotes its first and second derivatives w.r.t. its argument. 

\section{Preliminaries}
\vspace{-0.5mm}
We first state the formal definition and some preliminary results of the RDPF following \cite{blau:2019}. Subsequently, we introduce the $\alpha$-divergence measure used in the paper. 
\subsection{RDPF}
We first provide the mathematical definition of the RDPF for general alphabets.
\begin{definition}\textit{(RDPF)}\label{def:rdpf}
Let the information source $X$ be a random variable on $(\mathcal{X}, \mathbb{B}(\mathcal{X}))$ with distribution $\pdf{X} \in \mathcal{P}(\mathcal{X})$. 
Let $d: \mathcal{X} \times \mathcal{X} \to \mathbb{R}^+_0$ be a (measurable) distortion function satisfying $d(x,\hat{x})=0$ iff $x=\hat{x}$ and let $D: \mathcal{P}(\mathcal{X}) \times \mathcal{P}(\widehat{\mathcal{X}}) \to  \mathbb{R}^+_0$ be a divergence function with $D(p_X,p_{\hX})=0$ iff $p_X = p_{\hX}$. 
The RDPF is defined as:
\begin{align}
        R(D,P)\triangleq & \inf_{p_{\hX|X}}  I(X,\hX) \label{opt:RDPF} \\
        \textrm{s.t.}   & \quad \E{d(X,\hX)} \le D, \label{opt: distortion_constraint}\\
                        & \quad D(p_X||p_{\hX}) \le P \label{opt: perception_constraint} 
\end{align}
where the infimum is over all conditional distributions $p_{\hX|X}: \mathcal{X} \to \mathcal{P}(\mathcal{\hat{X}})$ and $I(X,\hX)$ denotes the mutual information \cite{Cover:2006} between the source $X$ and the reconstructed source $\hX$. 
\end{definition}
The following remark highlights the general properties of the RDPF.
\begin{remark}\textit{(On Definition \ref{def:rdpf})} The optimization problem \eqref{opt:RDPF} enjoys useful properties, under mild regularity conditions of the perception metric $D( \cdot\| \cdot)$.
In particular, \cite[Theorem 1]{blau:2019} shows that $R(D, P)$ is (i) monotonically non-increasing function in both arguments; (ii) convex if the divergence $D(\cdot||\cdot)$ is convex in its second argument. 
\end{remark}

\subsection{Alpha Divergence}
The $\alpha$-divergence $D_\alpha(p\|q)$ between two (normalized) probability distributions $p$ and $q$, using the Amari notation \cite{Amari:85}, is defined as 
\begin{align*}
    D_{\alpha_A}(p\|q) \triangleq \frac{4}{1-\alpha_A^2} \left(1 - \int p(x)^{\frac{1-\alpha_A}{2}} q(x)^{\frac{1+\alpha_A}{2}} \dx \right),
\end{align*}
for $\alpha_A \in \mathbb{R} \backslash \{\pm 1\}$.\footnote{The coefficient $4/(1-\alpha_A^2)$ guarantees that any $D_{\alpha_A}$ gives the same Fisher information metric when $p$ and $q$ are infinitesimally close.} The case with $\alpha_A=\pm 1$ is derived from limit $\alpha_A \to \pm 1$. Interestingly, $D_{\alpha_A}$ and $D_{-\alpha_A}$ are dual in the sense that $D_{\alpha_A}(p\|q) = D_{-\alpha_A}(q\|p)$ (reference duality).  

An alternate definition of $\alpha$-divergence between two (normalized) probability distributions $p$ and $q$ (see, e.g., \cite{Cressie:88}), which is used in the remainder of the paper, is given as follows:  
\begin{align}
    D_\alpha(p\|q) = \frac{1}{\alpha(\alpha - 1)} \left( \int_{-\infty}^{\infty} p(x)^{\alpha} q(x)^{1-\alpha} \,dx - 1 \right),
    \label{eq:Amari}
\end{align}
for $\alpha \in \mathbb{R} \backslash \{0, 1\}$. 
Since $\alpha = (1-\alpha_A)/2$, the reference duality is expressed by $D_{1-\alpha}(p\|q) = D_{\alpha}(q\|p)$. We also have $D_\alpha(p\|q) \geq 0$ with equality iff $p=q$, and that $D_\alpha(\cdot\|\cdot)$ is not symmetric (i.e., $D_\alpha(p\|q) \neq D_\alpha(q\|p)$) except for $\alpha = 0.5$. Furthermore, $D_\alpha(p\|q)$ is convex with respect to both arguments $p$ and $q$.\\

\subsubsection{The Role of $\alpha$} The value of the $\alpha$ parameter may have a significant effect on the resulting approximate distribution $q$. For $\alpha \leq 0$, the $\alpha$-divergence forces $q$ to have low density wherever $p$ has low density (zero-forcing). On the other hand, when $\alpha \geq 1$, the divergence is inclusive, i.e., it enforces $q>0$ wherever $p>0$, hence avoiding zero probability density in regions of the space in which $p$ has high density. When $\alpha \in (0, 1)$, the resulting $q$ distribution is intermediate between the two extreme possibilities. In particular, when $\alpha \to 0$, we should expect that the approximate distribution $q$ is more centered in the main mode of $p$ (mean seeking). In contrast, when $\alpha \to 1$, $q$ is expected to cover the target distribution $p$ more and capture more modes (mode-seeking).

\subsubsection{Special cases} Several widely used divergences are actually $\alpha$-divergences with specific values of $\alpha$. For instance, when $\alpha \to 1$ we obtain the KL divergence, when $\alpha \to 0$ we obtain the reverse KL divergence, when $\alpha \to -1$ we obtain the inverse Pearson divergence, when $\alpha = 1/2$ we obtain the Hellinger distance, and when $\alpha = 2$ the Pearson divergence. For $\alpha=0$, the $\alpha$-connection \cite{Amari:85} is the Levi-Civita connection under the Fisher-Rao metric. Note that the Rényi-$\alpha$ and the Tsallis-$\alpha$ divergences are closely related divergences (but not special cases). 

\subsubsection{Gaussian Case} Under the assumption that $p$ and $q$ belong to the set of Gaussian distributions, i.e., $p = \mathcal{N}(\mu, \sigma^2)$ and $q = \mathcal{N}(\nu, \rho^2)$, $D_{\alpha}(p\|q)$ can be analytically characterized as
\begin{align}
    \begin{split}
        D_\alpha(p\|q) &= \frac{1}{\alpha(1-\alpha)} \left( 1 - H_{\alpha}(p,q) \right)\\
        H_{\alpha}(p,q) &= \frac{\rho^\alpha\sigma^{1-\alpha}}{\sqrt{\alpha\rho^2+(1-\alpha)\sigma^2}}e^{-\frac{\alpha(1-\alpha)(\mu-\nu)^2}{2(\alpha\rho^2+(1-\alpha)\sigma^2)}}.
    \end{split} \label{eq:Alpha_Div_Gaussian}
\end{align}
However, \eqref{eq:Alpha_Div_Gaussian} is guaranteed to be real only when $\alpha \in [0, 1]$. For $\alpha \notin [0, 1]$, additional conditions must be imposed, to guarantee that $H_{\alpha}$ is a real-valued function, i.e., $\alpha\rho^2 + (1 - \alpha) \sigma^2 >0$, which ensures the existence of the integral in \eqref{eq:Amari}. 

For $\alpha > 1$, the condition becomes $\sigma^2 < \frac{\alpha}{\alpha-1}\rho^2$, indicating that the variance of $p$ should be smaller than the variance of $q$ multiplied by a constant factor. As $\alpha \to \infty$, this constant factor vanishes, and the constraint on the variances simplifies to $\sigma^2 \leq \rho^2$. On the other hand, as $\alpha \to 1^+$ the constraint disappears, i.e., $\sigma^2 < \infty$. For $\alpha < 0$, due to the duality of $\alpha$ divergence, we can apply the conclusions of the case $\alpha > 1$ by swapping the roles of $\sigma$ and $\rho$. The constraint shifts into $\sigma^2 > \frac{\alpha}{\alpha-1}\rho^2$. When $\alpha \to -\infty$, it becomes $\sigma^2 \geq \rho^2$ and when $\alpha \to 0^-$, the constraint disappears, $\sigma^2 > 0$.\vspace{-2mm}

\section{Main Results}
\vspace{-4mm}
In this section, we present our main results. Consider the RDPF problem in Definition \ref{def:rdpf} assuming that the information source is Gaussian distributed, i.e., $X \sim \mathcal{N}(\mu, \sigma^2)$, the perception measure \eqref{opt: perception_constraint} belongs to the family of $\alpha$-divergences, i.e., $D(\cdot||\cdot) = D_{\alpha}(\cdot||\cdot)$, and the distortion measure is the MSE metric, i.e., $d(X,\hat{X}) = (X - \hat{X})^2$. Following \cite{serra:2024}, we can characterize an upper bound to the RDPF for a Gaussian source $X$ by constraining the reconstruction distribution $\hat{X}$ to be jointly Gaussian with $X$ as follows.

\begin{definition} \textit{(JG-RDPF)}
    Let $D \ge 0$, $P \ge 0$. Then, assuming a Gaussian random variable $X$ is jointly Gaussian with a reconstruction variable $\hat{X} \sim q = \mathcal{N}( \nu, \rho^2)$, we define the joint Gaussian RDPF (JG-RDPF) $R^{G}(D,P)$ as follows:
    \begin{align}
        \begin{split}
        R^{G}(D,P) &= \min_{p_{\hX|X}, \nu, \rho^2}  \quad I(X,\hX) \\
        \text{s.t.} & \quad \mathbb{E}\left[(X - 
        \hat{X})^2\right] \leq D, \\
        & \quad D_\alpha(p\|q) \leq P.
        \end{split} \label{opt: RDPFG}
    \end{align}
\end{definition}

We remark that the definition $R^G(D, P)$ induces the following upper bound to $R(D, P)$. 
\begin{lemma}
    Let $D \ge 0$, $P \ge 0$. Then, $R(D,P)\leq R^G(D,P)$.
\end{lemma}
\begin{IEEEproof}
    The proof follows by observing that the constraint set of \eqref{opt: RDPFG} is a proper subset of the constraint set of \eqref{opt:RDPF}.
\end{IEEEproof}
We stress the following technical remark.
\begin{remark}
    The assumption of $\hat{X}$ being jointly Gaussian with $X$ leads to an upper bound to the optimal solution of the problem \eqref{opt: RDPFG}. This is because the distribution minimizing the perception constraint $D_\alpha(\cdot\|\cdot)$ cannot be guaranteed to be Gaussian. However, in the case of reverse KL divergence, i.e., ($\alpha \to 0$), where it is proven that the minimizing distribution is itself Gaussian \cite{Fang:2020}, the above upper bound is exact.
\end{remark}

We note that, without loss of optimality, in \eqref{opt: RDPFG} we can assume that $\mu = \nu = 0$ since this does not affect the mutual information $I(X,\hat{X})$ while being optimal in terms of MSE and $D_{\alpha}$ metrics, as shown in the following lemma.

\begin{lemma}
     Let $\Delta = \mu - \nu$. Then, $D_\alpha(p\|q)$ has a global minimum at $\Delta = 0$. 
\end{lemma}
\begin{IEEEproof}
    For $\alpha \in (-\infty,0) \cup (1,+\infty)$, we see that $D_{\alpha}(p\|q) \approx s(\Delta) = e^{\Delta^2}-c$, where $c$ denotes constant terms. The first and second derivative of $s(\Delta)$ are, respectively, 
    \begin{align*}
        s'(\Delta) &= 2 \Delta e^{\Delta^2} , \qquad
        s''(\Delta) &= 2e^{\Delta^2}+4\Delta^2e^{\Delta^2} > 0.
    \end{align*}
    We conclude that $s(\Delta)$ is strongly convex and therefore minimized by its stationary point $\Delta_0 = 0$. Hence, also $D_{\alpha}(p\|q)$ has a global minimum at $\Delta_0 = 0$. 
    
    For $\alpha \in (0,1)$,  $D_{\alpha}(p,q) \approx t(\Delta) = c-e^{-\Delta^2}$. The first and second derivative of $t(\Delta)$ are, respectively,
    \begin{align*}
        t'(\Delta) = 2\Delta e^{-\Delta^2}, \qquad t''(\Delta) = 2e^{-\Delta^2}(1-2\Delta^2),
    \end{align*}
    from which we characterize the stationary point $\Delta_0 = 0$. For $\Delta < 0$, $t'(\Delta)$ is negative which means that $t(\Delta)$ is decreasing where for $\Delta > 0$, $t'(\Delta)$ is positive meaning that $t(\Delta)$ is increasing. For $\Delta \in (-\infty,-\frac{1}{\sqrt{2}}) \cup (\frac{1}{\sqrt{2}},+\infty)$, $t''(\Delta)$ is negative where for $\Delta \in (-\frac{1}{\sqrt{2}},\frac{1}{\sqrt{2}})$ is positive. Therefore, the function $t(\Delta)$ and thus $D_{\alpha}(p\|q)$ has a global minimum at $\Delta_0 = 0$. This concludes the proof.
\end{IEEEproof}

The solution of \eqref{opt: RDPFG} can be greatly facilitated by an in-depth analysis of the perception constraint $D_{\alpha}(p\|q) \le P$, focusing in particular on the case where it holds with equality. The following lemma characterizes essential aspects of this case.

\begin{lemma} \label{lemma: char_exp_pol}
    Assume $P \in [0,P_{\alpha,\max}]$ and let $p = \mathcal{N}(0, \sigma^2)$ and $q = \mathcal{N}(0, \rho^2)$. Then the constraint $D_{\alpha}(p\|q) = P$ can be expressed as $f(x) = 0$, where
    \begin{align}
            f(x) = x^{\alpha} - \alpha C x - (1 - \alpha)C\label{eq: alpha_exp_pol}
    \end{align}
    with $x = \frac{\rho^2}{\sigma^2}$ and $C = (1 - \alpha(1-\alpha)P)^2$. Furthermore, for all $\alpha \in (-\infty, 0) \cup (0,1) \cup (1, +\infty)$, \eqref{eq: alpha_exp_pol} has two roots, $r_0$ and $r_1$, such that $r_0 \in [0, x_0]$ and $r_1 \in [x_0, y_0]$, where $x_0$ is the unique stationary point of \eqref{eq: alpha_exp_pol} and $y_0 = (x_0 + \epsilon) - \frac{f(x_0 + \epsilon)}{f'(x_0 + \epsilon)}$ for $\epsilon > 0$.
\end{lemma}
\begin{IEEEproof}
    The equivalence between  $D_{\alpha}(p\|q) = P$ and \eqref{eq: alpha_exp_pol} easily follows from algebraic manipulations.  
    We examine the properties of equation \eqref{eq: alpha_exp_pol} through its derivatives, i.e, 
    \begin{align*}
        f'(x) &= \alpha (x^{\alpha-1} - C), \qquad     f''(x) = \alpha\left(1-\alpha\right)x^{\alpha-2}. 
    \end{align*}
    Considering $f'(x)$, we can characterize the stationary point $x_0$ of $f(x)$, i.e., $f'(x) = 0\overset{\alpha\neq 0,1}{\Rightarrow} x_0 = C^{\frac{1}{\alpha-1}}$. For \( \alpha \in (-\infty, 0) \cup (1, +\infty) \), \( f''(x) \geq 0 \), implying that \(f'(x)\) is increasing and \(f(x)\) is concave up. Therefore, \(x_0\) will be the global minimum of \(f(x)\). For \( \alpha \in (0, 1) \), it is \( f''(x) \leq 0 \) which means \(f'(x)\) is decreasing and \(f(x)\) is concave down. Hence, \(x_0\) will be the global maximum of \(f(x)\). We can now infer about the number of roots of $f(x)$ by determining the sign of \(f(x_0)\), i.e.,
\begin{align*}
    f(x_0) = \left(1-\alpha\right)C\left(C^{\frac{1}{\alpha-1}}-1\right).
\end{align*}
By definition, the condition $C \geq 0$ holds. We notice that for $C = 0$, $x_0 = 0$ is the unique root of the function. 
To evaluate the sign of $f(x_0)$, we can verify that the following holds
\begin{align*}
   \alpha < 0 \Rightarrow C^{\frac{1}{\alpha-1}}-1 \leq 0 \Rightarrow f(x_0) \leq 0,\\
   0 < \alpha < 1 \Rightarrow
    C^{\frac{1}{\alpha-1}}-1 \ge 0 \Rightarrow f(x_0) \geq 0,\\
   \alpha > 1 \Rightarrow C^{\frac{1}{\alpha-1}}-1 \ge 0 \Rightarrow f(x_0) \leq 0.
\end{align*}

This implies that when $f(x_0) \le 0$ (resp. $f(x_0) \ge 0$) we have a positive (resp. negative) second derivative. Since $f'(x)$ is monotonic, $f(0) = -(1 - \alpha)C$ and 
\begin{align*}
    \lim_{x \to +\infty} f(x) &= \begin{cases}
        + \infty ~~ \text{if} ~~ \alpha \in (-\infty, 0) \cup (1, +\infty)\\
        - \infty ~~ \text{if} ~~ \alpha \in (0,1)\\
    \end{cases}
\end{align*}
we can deduct that $f(x)$ necessarily has two roots in $\mathbb{R}$.
\par We now identify the sets where the roots $r_0$ and $r_1$ reside. Starting with $r_0$, we can notice that $f(x)$ is monotone on the set $[0,x_0]$ while changing the sign in its extremes. Hence, the root $r_0$ must belong to this set.
In the case of $r_1$, we notice that, for $\epsilon > 0$, the following inequalities hold
\begin{align*}
    &\begin{cases}
        f(y) \ge f_T(y) ~~ \text{if} ~~\alpha \in (-\infty, 0) \cup (1, +\infty)\\
        f(y) \le f_T(y) ~~ \text{if} ~~\alpha \in (0,1)
    \end{cases}\\
    &f_T(y) = f(x_0 + \epsilon) + f'(x_0 + \epsilon)(y - (x_0 + \epsilon))
\end{align*}
due to the convexity (resp. concavity) of $f$. This allows us to characterize the set where $r_1$ resides by taking as inferior $x_0$ and as superior the zero of the linear function $f_T(y)$, i.e., $y_0 = (x_0 + \epsilon) - f(x_0 + \epsilon)(f'(x_0 + \epsilon))^{-1}$. This concludes the proof.
\end{IEEEproof}
We stress the following point regarding the evaluation of the roots $r_0$ and $r_1$.
\begin{remark}
    Lemma \ref{lemma: char_exp_pol} proves that, in their respective interval, $r_0$ and $r_1$ are the unique roots of \eqref{eq: alpha_exp_pol}. This property facilitates the numerical estimation of the problem by allowing the bisection method to be applied directly \cite[Chapter 2.1]{burden:2015} in each interval.
\end{remark}

Equipped with the previous lemma, we are ready to develop a parametric solution for the JG-RDPF in \eqref{opt: RDPFG}.

\begin{theorem} \textit{(Parametric JG-RDPF solution)}\label{thm:gaussian_rdp}
Let $X$ be a scalar Gaussian source $X \sim \mathcal{N}(0,\sigma^2)$. Then, the JG-RDPF $R^G(D,P)$ under squared error distortion and $\alpha$-divergence perception is achieved by a jointly Gaussian reconstruction $\hat{X} \sim \mathcal{N}(0,\rho^2)$ and is given by
\begin{align}
    R^G(D, P)
    &= \begin{cases}
        \max\left\{\frac{1}{2}\log\frac{\sigma^2}{D}, 0\right\} 
            &\quad\textit{if } (D,P) \in \mathcal{S} \\ 
        \frac{1}{2}\log\frac{2 \rho^2\sigma^2}{\rho^2\sigma^2 - \left( \frac{\sigma^2 + \rho^2 - D}{2} \right)^2}  &\quad\textit{if } (D,P) \notin \mathcal{S}.
    \end{cases}\label{eq: RDPF_alpha_closed}
\end{align}
where
\begin{align*}
    \rho^2
    &= \begin{cases}
        \sigma^2 - D &\quad\textit{if } (D,P) \in \mathcal{S} \\ 
        \min\{r_0,r_1\}  &\quad\textit{if } (D,P) \notin \mathcal{S}. 
    \end{cases} \nonumber\\
    \mathcal{S} &= \Big\{(D,P) \in \mathbb{R_+}^2: P > g(D,\sigma)~~\land~~\\
    & \qquad (\alpha - 1)\left(\left|1 - \frac{D}{\sigma^2}\right| - \left(1 - \frac{1}{\alpha}\right) \right) > 0 \Big\}\\
    g(D,\sigma) &= \frac{1}{\alpha(1-\alpha)}\left(1-\frac{\sigma^{1-\alpha}|\sigma^2-D|^{\alpha/2}}{\sqrt{\alpha|\sigma^2-D|+(1-\alpha)\sigma^2}}\right),
\end{align*}
with $r_0$ and $r_1$ being the roots of \eqref{eq: alpha_exp_pol}.
\end{theorem}
\begin{IEEEproof}
The objective function and constraints of \eqref{opt: RDPFG} can be rewritten in terms of parameters of the Gaussian source $X \sim \mathcal{N}(0,\sigma^2)$ and reconstruction $\hat{X}\sim \mathcal{N}(0,\rho^2)$ as follows
\begin{align}
    I(X,\hat{X}) &= \frac{1}{2}\log\frac{\rho^2\sigma^2}{\rho^2\sigma^2-\theta^2}, \label{eq: MI_th1}\\
    \mathbb{E}\left[\left(X-\hat{X}\right)^2\right] &= \sigma^2+\rho^2-2\theta \leq D,\label{eq: RDPF_dis_th1}\\
    D_\alpha(p\|q) = \frac{1}{\alpha(1-\alpha)} \Big( 1& - \frac{\rho^\alpha\sigma^{1-\alpha}}{\sqrt{\alpha\rho^2+(1-\alpha)\sigma^2}}\Big) \leq P \label{eq: RDPF_div_th1}
\end{align}
where $\theta = \mathbb{E}[X\hat{X}]$ is the covariance between $X$ and $\hat{X}$.

We derive the closed-form solution of $R^G(D,P)$ by considering the different cases when only the distortion constraint is active (\textbf{Case I}), when only the perception constraint is active (\textbf{Case II}), or when both are active (\textbf{Case III}). 
\par \textbf{Case I:} When the perception constraint is not active (i.e., \( P = \infty \)), the solution to \eqref{opt: RDPFG} reduces to the rate-distortion function of a Gaussian source \cite[Theorem 10.3.2]{Cover:2006} 
\begin{align*}
    R^G(D,\infty) =
    \begin{cases}
        \frac{1}{2} \log\frac{\sigma^2}{D} & 0 \leq D \leq \sigma^2 \\
        0 & D > \sigma^2
    \end{cases},
\end{align*}
which is attained by some $p_{\hX|X}$ with marginal distribution $\hat{X} \sim \mathcal{N}(0, \sigma^2 - D)$. Furthermore, since the distortion constraint is active and holds with equality while the perception constraint is inactive, $\rho = \sqrt{\sigma^2-D}  < \sigma$ and \eqref{eq: RDPF_div_th1}
must hold with strict inequality. However, we note that for $\alpha > 1$ the argument of the root in \eqref{eq: RDPF_div_th1} may be negative, hence the constraint can become ill-posed. To avoid this case, we enforce the following additional constraint
\begin{align*}
    \left|1 - \frac{D}{\sigma^2}\right| > \left(1 - \frac{1}{\alpha}\right).
\end{align*}
\par \textbf{Case II:} When only the perception constraint is active, we can characterize the variance of the reconstruction $\rho^2$ by solving the generalized equation \eqref{eq: alpha_exp_pol},  following the results of Lemma \ref{lemma: char_exp_pol}. Furthermore, the inactive distortion constraint implies that \eqref{eq: RDPF_dis_th1} holds with strict inequality for all $\theta \in [0, \sigma\rho]$. We can therefore set $\theta = 0$, which in turns implies that $R^G(\infty, P) = 0$.

\par \textbf{Case III:} When both constraints hold with equality, the resulting system of equations can be used to characterize both the variance $\rho^2$ and the covariance $\theta$. Using Lemma \ref{lemma: char_exp_pol}, we can find the two roots of \eqref{eq: alpha_exp_pol}, i.e., $\rho^2_0$ and $\rho^2_1$. Despite both roots being optimal solutions for \eqref{eq: alpha_exp_pol}, we select $\rho^2 = \min\{\rho_0^2, \rho_1^2\}$ under the argument that a lower variance of the reconstruction is optimal in terms of MSE distortion. From the equality of the distortion constraint, we can instead derive $\theta = \tfrac{1}{2}(\sigma^2 + \rho^2 - D)$.
which, once substituted in \eqref{eq: MI_th1}, recovers the closed-form expression for $R^G(D,P)$ in \eqref{eq: RDPF_alpha_closed}. This concludes the proof.
\end{IEEEproof}

\par \textit{JG-RDPF Special Cases:} We shall notice that for specific instances of $D_\alpha$ we can directly recover previously characterized analytical JG-RDPF \cite{serra:2024}, i.e., Hellinger distance ($\alpha = \tfrac{1}{2}$), direct and reverse Kullback-Leibler divergence ($\alpha \to 1$ and $\alpha \to 0$, respectively). In the remainder of this section, we derive a new characterization for the Pearson divergence ($\alpha = 2$).
We should mention that in each of these cases, the roots of the generalized polynomial \eqref{eq: alpha_exp_pol} can be easily found analytically, allowing for the closed-form characterization of $\rho(\sigma,P)$, based on which we recover the closed-form expression of $R^G(D,P)$ once substituted in \eqref{eq: RDPF_alpha_closed}. 

\textit{Pearson Divergence} (\(\alpha=2\)):
    
        \((\ref{eq: alpha_exp_pol}) \overset{\alpha=2}{\Rightarrow}\rho^4 - 2(1+2P)^2\rho^2\sigma^2 + (1+2P)^2\sigma^4 = 0\)
        \[
            \rho^2 = 
            \begin{cases}
            \begin{aligned}
            &\sigma^2\left(1 + 2P\right) \left(1+2P+2\sqrt{P+P^2} \right) \\
            &\qquad\text{if } \rho^2-\left(1+2P\right)^2\sigma^2>0,
            \end{aligned} \\
            \begin{aligned}
            &\sigma^2\left(1 + 2P\right) \left(1+2P-2\sqrt{P+P^2} \right) \\
            &\qquad\text{if } \rho^2-\left(1+2P\right)^2\sigma^2\leq0.
            \end{aligned}
            \end{cases}
        \]
        
    
\section{Numerical Results}
In this section, we provide plots that visually illustrate the main results and provide insights into the behavior and functionality of the RDPF.
In all the presented examples, we consider a Gaussian source $X \sim \mathcal{N}(0, 1)$.

\begin{figure}[htbp]
    \centerline{\includegraphics[width=0.35\textwidth]{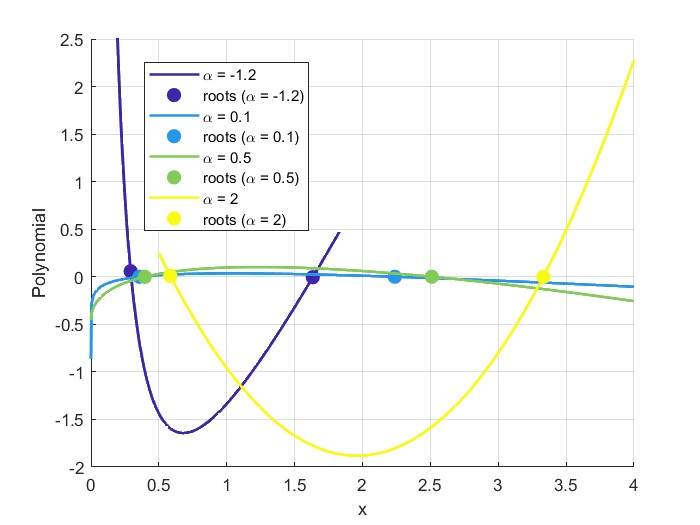}}
    \caption{Polynomial of \eqref{eq: alpha_exp_pol} for  $\alpha = -1.2$, $\alpha = 0.1$, $\alpha = 0.5$ and $\alpha = 2$, with the perception constraint fixed to $P = 0.2$.}
    \label{fig:polynomial}
\end{figure}

\begin{figure}[t]
    \centerline{\includegraphics[width=0.5\textwidth]{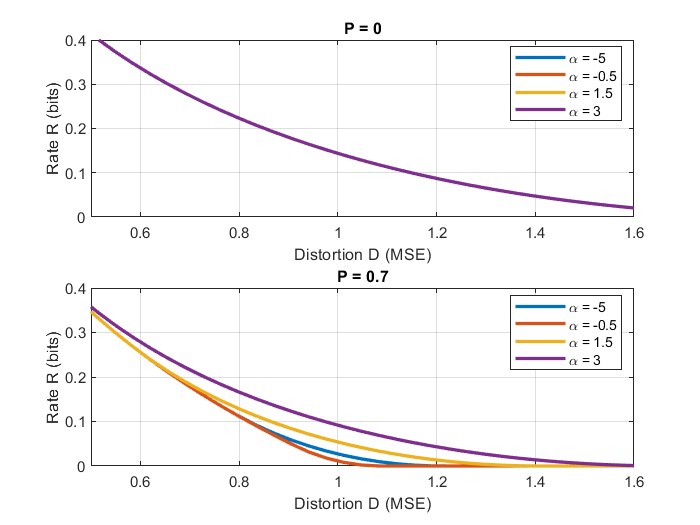}}
    \caption{Rate-distortion curves for $\alpha = -5$, $\alpha = -0.5$, $\alpha = 1.5$ and $\alpha = 3$. The first figure represents $P = 0$, whereas the second $P=0.7$.}
    \label{fig:RD-various-alpha}
\end{figure}

\begin{figure}[htbp]
    \centerline{\includegraphics[width=8.5cm]{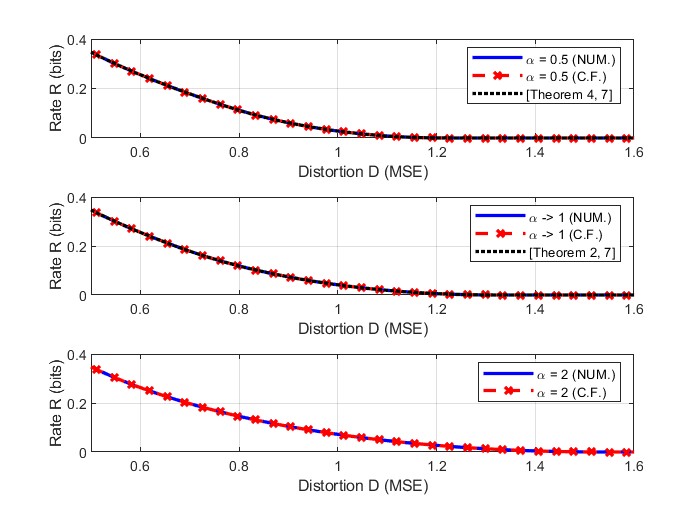}}
    \caption{Rate-distortion curves for the special cases of $\alpha = 0.5$, $\alpha \to 1$ and $\alpha = 2$. The perception constraint is fixed $P=0.5$. }
    \label{fig:RD-converges}
\end{figure}

\paragraph{Polynomial representation of perception constraint}
Fig.~\ref{fig:polynomial} plots the polynomial \eqref{eq: alpha_exp_pol} for different values of $\alpha$ and fixed perception constraint $P = 0.2$, taking into account the constraint of the variances $\alpha\rho^2 + (1 - \alpha) \sigma^2 >0$. We see that for every value of $\alpha$, the polynomial has two roots in the disjoint sets defined based on the global extrema.
\paragraph{Comparison between $\alpha$-divergences}
Fig.~\ref{fig:RD-various-alpha} plots the RD curve for varying $\alpha$ values. As expected, we see that for $P=0$ (perfect perceptual quality), all curves overlap. On the other hand, when $P=0.7$, the curves diverge from each other, since different $\alpha$ values lead to different perception measures. We notice that as $\alpha$ increases, the curves start to differ more from each other indicating that the perceptual constraint becomes more stringent, leading to higher distortion for the same rate compared to lower $\alpha$ values. 
Fig.~\ref{fig:RD-converges} evinces the matching between our analytical closed-form solutions and our numerical results for the special cases of $\alpha = 0.5$, $\alpha \to 1$ and $\alpha = 2$. In the first two cases, we confirm our results with \cite{serra:2024}, Theorems 4 and 2, which provide a closed-form solution for JG-RDPF with Hellinger and KL divergence as perception constraints, respectively.

\section*{Acknowledgment}
This work is part of a project that has received funding from the European Research Council (ERC) under the EU’s Horizon 2020 research and innovation programme (Grant agreement No. 101003431), from which the work of M. Kountouris and G. Serra is supported. The work of M. Sourla was supported by the ERASMUS+ programme and the work of P. A. Stavrou is supported by the SNS JU project 6G-GOALS \cite{strinati:2024} under the EU’s Horizon programme Grant Agreement No. 101139232.

\bibliographystyle{IEEEtran}
\bibliography{references}

\end{document}